%% file: main.tex
\documentclass[runningheads]{llncs}

\usepackage{preamble}
\usepackage{graphicx}
\begin{document}
%
\title{Multi-scale MRI reconstruction via dilated ensemble networks}
\author{Wendi Ma\inst{1}\and
Marlon Bran Lorenzana\inst{1}\and
Wei Dai\inst{1}\and
Hongfu Sun\inst{1}\and
Shekhar S. Chandra\inst{1}}
%
%
\institute{The University of Queensland\\
}
%
\maketitle              
\begin{abstract}
As aliasing artefacts are highly structural and non-local, many MRI reconstruction networks use pooling to enlarge filter coverage and incorporate global context. However, this inadvertently impedes fine detail recovery as downsampling creates a resolution bottleneck. Moreover, real and imaginary features are commonly split into separate channels, discarding phase information particularly important to high frequency textures. In this work, we introduce an efficient multi-scale reconstruction network using dilated convolutions to preserve resolution and experiment with a complex-valued version using complex convolutions. Inspired by parallel dilated filters, multiple receptive fields are processed simultaneously with branches that see both large structural artefacts and fine local features. We also adopt dense residual connections for feature aggregation to efficiently increase scale and the deep cascade global architecture to reduce overfitting. The real-valued version of this model outperformed common reconstruction architectures as well as a state-of-the-art multi-scale network whilst being three times more efficient. The complex-valued network yielded better qualitative results when more phase information was present.

\end{abstract}
\section{Introduction}
Magnetic resonance imaging (MRI) is a widely used clinical diagnostic technique for capturing soft-tissue and organ anatomy. As its sequential acquisition of $k$-space  is quite slow, undersampling is often used to save time but it violates the Nyquist-Shannon sampling theorem and introduces unwanted aliasing artefacts. However, \cite{tao} showed that given sufficient a priori information, one can recover the majority of a signal even when sampled below the Nyquist rate. In recent years, it has become common to use deep neural networks to reconstruct and denoise undersampled MR images by leveraging existing fully-sampled training data.

Although $k$-space is inherently complex-valued, many works still extract magnitude-only images or otherwise stack real and imaginary components as separate features. In this way, they cannot perform true complex valued convolution and phase information is discarded. \cite{deepcomplexnetworks} developed building blocks for complex-valued CNNs which a few works \cite{complex_unet}\cite{analysisofcomplexnetworks} have observed offer improved reconstruction quality.

Recent works \cite{deeplab,dudornet} have also emphasized the importance of receptive fields in vision tasks, wherein larger receptive fields allow a model to capture more global context in addition to local features of images. As standard convolution operations only linearly enlarge receptive fields \cite{receptive_fields}, traditional CNNs like U-Net reconstruction models rely on successive pooling layers to target global features. However, since downsampling reduces resolution, \cite{dilated} used dilated convolutions in lieu of pooling to preserve resolution and better capture fine details. Motivated by the idea that image features occur across a wide range of scales, \cite{neuralphotoediting} designed a `multiscaled dilated convolution block' inspired by the famous Inception module \cite{inception}. Like it, it simultaneously processes different features scales but uses dilated rather than different sized filters, which is more expensive. In similar fashion, DeepLab's Atrous Spatial Pyramid Pooling (ASPP) layers uses parallel dilated filters to extract and combine multi-scale features \cite{deeplab} . Dilated convolutions have also been used in MRI reconstruction by DuDorNet \cite{dudornet}. However, it utilized dense residual connections to aggregate multi-scale features between dilated convolution layers rather than a parallelized approach.

Inspired by these prior works, we develop a multi-scale reconstruction network comprising 1.) an `Ensemble Denoiser Block' (Figure \ref{fig:ensemble}) using parallel denoiser branches to target different feature scales. Each branch uses a densely connected structure similar to \cite{dudornet} but each with a different combination of dilation rates to achieve different RF sizes. 2.) a global architecture based on the deep cascade architecture presented in \cite{deepcascade}. We cascade together Ensemble Denoiser Blocks and also introduce additional dense connections between them.

\section{Methods}
\subsection{Problem Formulation}
Let $k_f$ be a fully-sampled $k$-space and $x_f \defeq \fourier^{-1}(k_f)$ a fully-sampled image reconstructed from it. Given a binary sampling mask $M$, we can use it to obtain an undersampled $k$-space $k_u$ and undersampled image $x_u$ as per \eqref{eq:undersampling}. Note that $\fourier$ and $\fourier^{-1}$ denote the Fourier and Inverse Fourier Transforms and $\odot$ the Hadamard product.
\begin{align}
    k_u = M \odot k_f \implies x_u = \fourier^{-1}\br{k_u} = \fourier^{-1}\br{M \odot \fourier \br{x_f}}
    \label{eq:undersampling}
\end{align}
We wish to train a reconstruction network $f$ to predict the value of $x_f$ given $x_u$ and $M$. To train it, the following two-part optimization problem was posed
\begin{equation}
    \argmin_{\theta} \br{\norm{x_f - f\br{x_u; \theta}}_1 + \norm{k_u - M \odot \fourier\br{f\br{x_u; \theta}}}_1} \label{eq:problem}
\end{equation}
consisting of $\mathcal{L}_1$ loss between $x_f$ and the network's output predictions, as well as a data-consistency constraint like that in \cite{deepcascade}. The latter ensures the preservation of the originally sampled $k$-space points in $k_u$.

\subsection{Complex-valued Networks}
For the complex-valued version of our network, we use the complex building blocks introduced by \cite{deepcomplexnetworks}, particularly their implementation of complex convolution, complex batch-normalization and $\crelu$ activation.
The 2D complex convolution layer exploits distributivity to decompose the operation into four separate real-valued convolutions. For example, given a complex input $\bm{h} = \bm{a} + i\bm{b}$ and complex kernel $\bm{W} = \bm{W_R} + i\bm{W_I}$, we obtain
\begin{equation}\label{eq:complex_conv} 
        \citeequation{\bm{W} * \bm{h} = \br{\bm{A}*\bm{x} - \bm{B}*\bm{y}} + i\br{\bm{B}*\bm{x} + \bm{A}*\bm{y}}}{deepcomplexnetworks}
\end{equation}

As regular batch normalization does not ensure equal variance across both real and imaginary components, \cite{deepcomplexnetworks} formulated a version using vector whitening to standardize inputs to the complex Gaussian distribution. Finally, $\crelu$ simply applies ReLU on the real and imaginary parts of an input separately and sums them. It exhibited the best convergence characteristics and phase information of the complex activations tested in \cite{deepcomplexnetworks}.

\subsection{Receptive Fields and Dilated Convolutions}
The `receptive field' of a CNN is defined as the region of an input image which effects the value of each individual feature/pixel in the output \cite{receptive_fields}. Larger receptive fields are typically desirable as each output feature `sees' more of the input and so can take into account more global context. Dilated convolutions are able to achieve arbitrary receptive field sizes (RF sizes) with minimal resolution loss. Whereas regular convolution applies its filters to a contiguous image patch, dilated convolution introduces gaps into the kernels so that adjacent weights of the filter are applied to image pixels a fixed distance apart. This distance is known as the dilation rate and expands the filter's effective RF size in accordance to \eqref{eq:rf_recursive}. Of course, it is still unreasonable to try to achieve very large receptive fields in a single convolution - the sparsity of pixels sampled by each filter would itself cause a great deal of information loss. Instead, it is common to exponentially grow receptive fields over a small number of sequential layers with reasonable dilation rates \cite{dautomap,dudornet}. \cite{receptive_fields} developed the following recurrence relation and corresponding closed-form expression for sequential layers
\begin{equation}\label{eq:rf_recursive}
    \citeequation{r_{l-1} = r_l + \alpha_l(k_l - 1), \quad r_0 = \sum_{l=1}^{L} \br{\alpha_l\br{k_l - 1}} + 1}{receptive_fields}
\end{equation}
where $r_l$ denotes the RF size of each layer, $k_l$ the kernel size, $\alpha_l$ the dilation rate and assuming a stride of $1$.

\newpage
\subsection{Ensemble Denoiser Block}
\begin{figure}
    \centering
\includegraphics[width=0.7\textwidth]{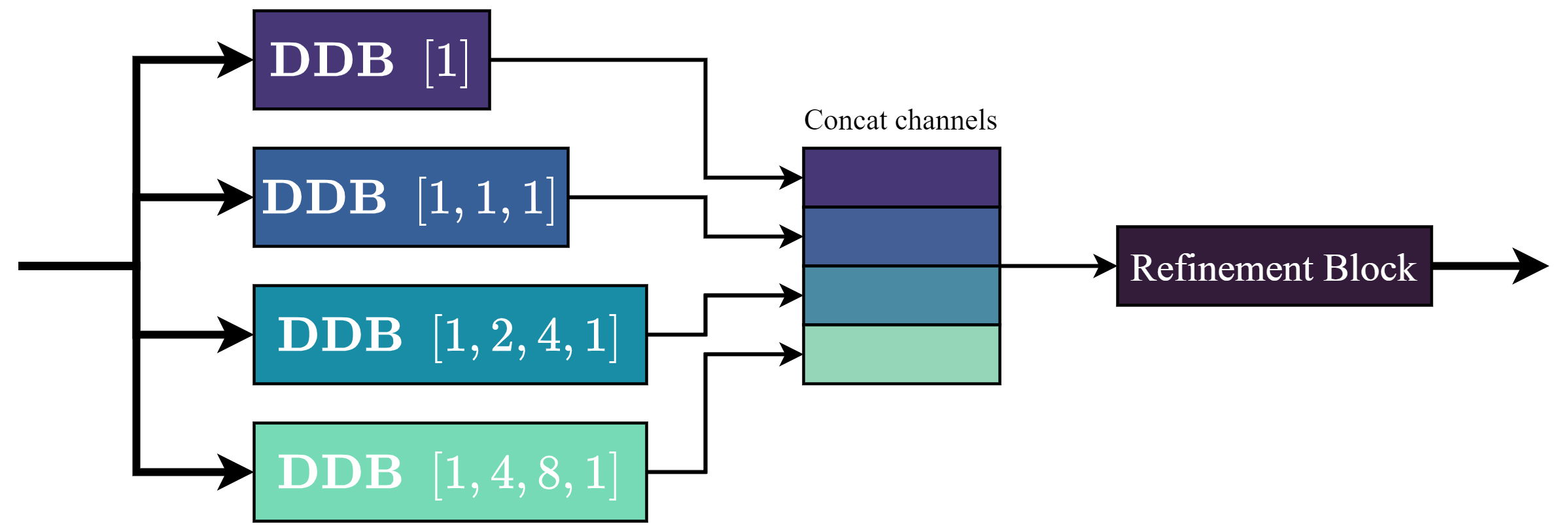}
    \caption{The architecture of our proposed Ensemble Denoiser Block comprising parallel branches targetting different feature-scales  DDB denotes the Dense Dilated Block from Figure \ref{fig:ddb} with dilation factors denoted in square brackets.}
    \label{fig:ensemble}
\end{figure}
\begin{figure}[ht]
    \centering
    \includegraphics[width=1\textwidth]{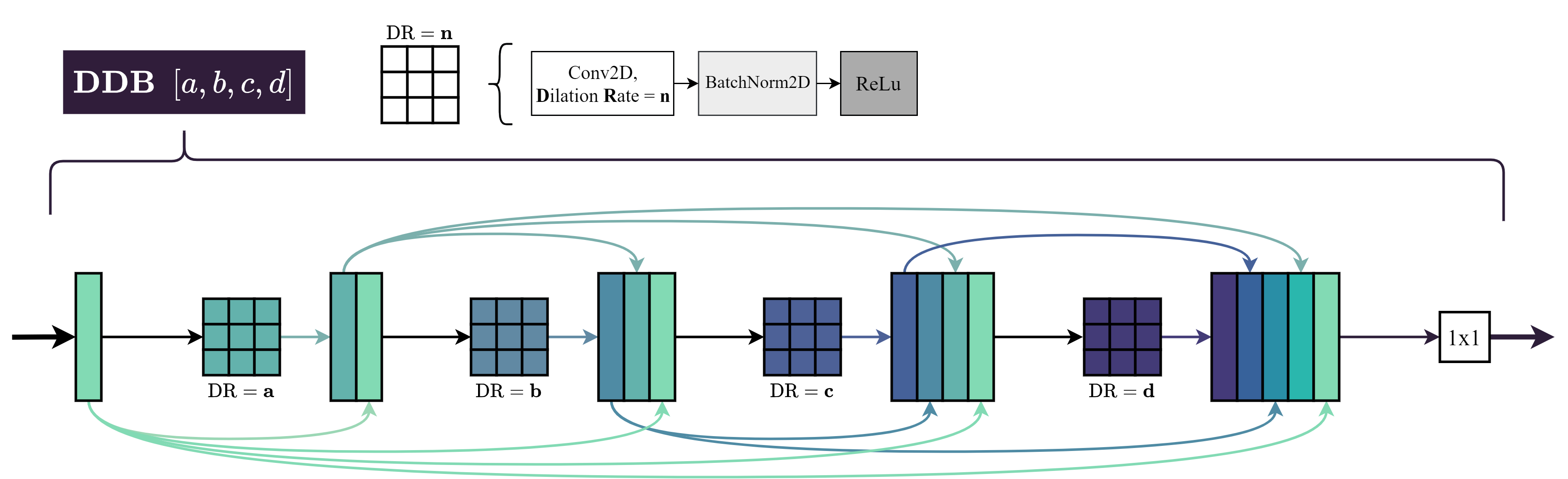}
    \caption{The architecture of the Dilated Dense Block (DDB) comprising dilated convolution blocks and dense residual connections.}
    \label{fig:ddb}
\end{figure}

We propose an Ensemble Denoiser Block (Figure \ref{fig:ensemble}) which aims to capture multi-scale features simultaneously using an ensemble approach. Input images are fed into parallel CNN denoiser branches whose respective outputs are then combined to produce a higher fidelity reconstruction. Like the parallel filters in previous works \cite{deeplab,neuralphotoediting}, we construct each branch to have different overall RF sizes, in our case $3\times3, 7\times7, 17\times17$ and $35\times35$ respectively. Intuitively, this will force them to focus on different feature scales, i.e. small receptive fields should encourage the restoration of fine detail whilst larger ones larger structures and global anatomy. This was supported by initial visualizations on a toy dataset (Figure \ref{eq:rf_recursive}) in which branches first appeared to reproduce textures and tones and then coarser shapes and outlines as we enlarged the receptive fields. Moreover, the parallel computation of these branches should also be more efficient than the SDRDB blocks in \cite{dudornet} which were connected sequentially.
\begin{figure}[ht]
    \centering
    \includegraphics[width=0.6\textwidth]{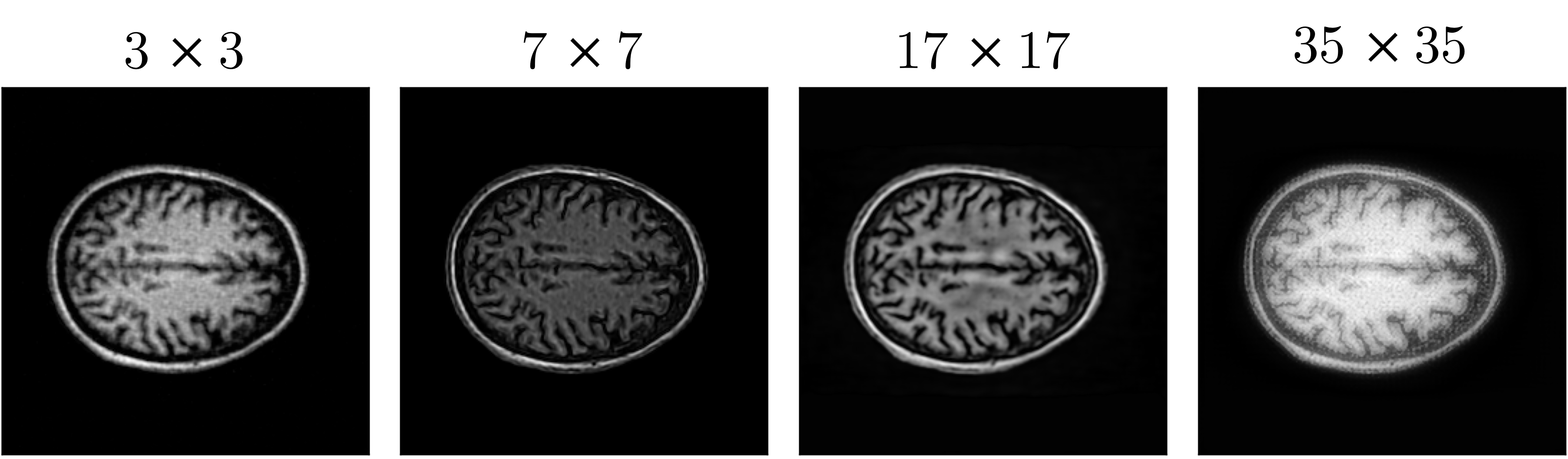}
    \caption{Intermediate branch outputs visualized on the OASIS dataset \cite{oasis}. Smaller RF sizes reproduce local features (textures, tones) and larger sizes global features (shapes, outlines).}
    \label{fig:learner_outputs}
\end{figure}
The branches themselves (Figure \ref{fig:ddb}) comprise dense blocks \cite{dense} with layer depths and dilation rates chosen to attain their chosen RF sizes. Like \cite{dudornet}, we chose this architecture to add additional scale by combining feature maps from many different RF sizes. Since intermediate feature maps are accessible by all subsequent layers, dense connections also encourage feature reuse and enable model compression as well as improved gradient propagation and learning \cite{deepcascade}.
Between consecutive convolution layers, 2D batch normalization is performed and non-linearity introduced through ReLu activation. After each branch, a 1x1 convolution is used to obtain a single-channel intermediate reconstruction and then the `refinement block' from \cite{dautomap} used to combine each branch's reconstruction. This contains two convolution layers for feature extraction followed by a transpose convolution layer to build the final output image.

\subsection{Global Structure}
\begin{figure}[ht]
    \centering
    \includegraphics[width=1\textwidth]{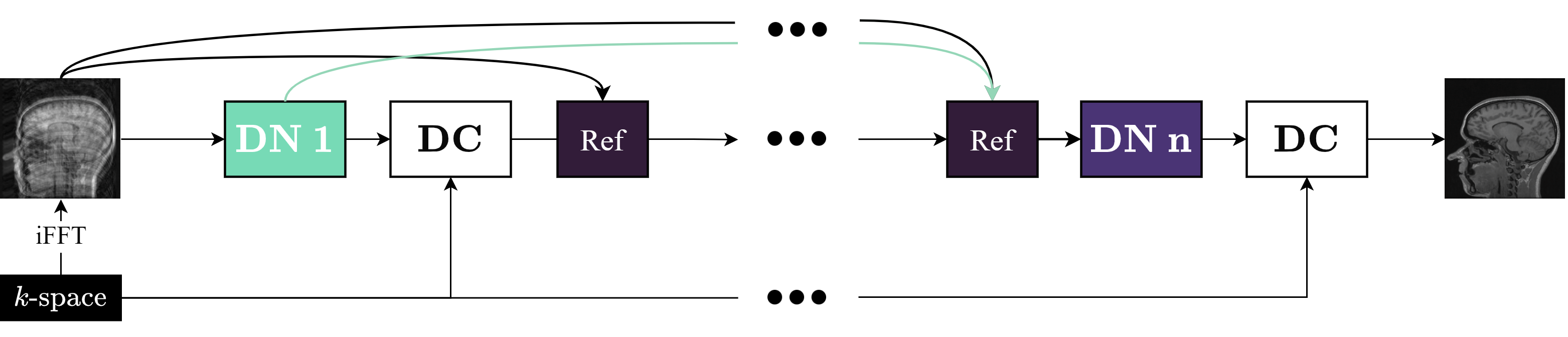}
    \caption{The general structure for an $n$ length cascade network. DN represents an Ensemble Denoiser Network, DC data-consistency and Ref a refinement operation which channel-wise concatenates its inputs and feeds them through a refinement block.}
    \label{fig:cascades}
\end{figure}

Many reconstruction works have adopted \cite{deepcascade}'s deep cascade architecture, wherein multiple denoising networks are connected together with interleaved data consistency operations. Data-consistency enforces the homonymous constraint in \eqref{eq:problem} by directly restoring already sampled $k$-space values to the output of each cascaded network which encourages networks to focus only on recovering missing $k$-space features. The cascade approach transforms a single image reconstruction task into a series of end-to-end reconstruction problems and has been shown to less prone to overfitting than directly increasing network depth \cite{deepcascade}.
In our network (Figure \ref{fig:cascades}), five Ensemble Denoiser Blocks are cascaded together to form the overall network. Dense connections were also added between each Ensemble Denoiser Block to try combine the best features from intermediate reconstructions.

\subsection{Experimental Configuration}
In our experiments, we use subsets of the Calgary-Campinas-359 (CC-359) \cite{calgary} and FastMRI \cite{fastmri} datasets. For CC-359, we used the single-channel coil portion constituting 45 fully-sampled T1-weighted brain scans with a 25-10-10 split for training, validation and testing. The slices were given as complex $256\times256$ images but contained minimal phase information. For FastMRI, we took a subset of the single-coil knee scans comprising 7135 raw 2D k-space slices with full phase information. These were subject to iFFT, cropped to $320\times320$ complex images and distributed per a 60\%-20\%-20\% data split. We also performed magnitude normalization on both datasets and emulated undersampling using 1D Gaussian masks at reduction factors of 4, 6 and 8 with corresponding centre-tile sizes of \sfrac{1}{20}, \sfrac{1}{40} and \sfrac{1}{80} times image width.

We trained two version of our models, the complex-valued $\mathbb{C}$-C5ED and its real-chanelled counterpart C5ED (Cascade 5 Ensemble Denoiser). We also trained the complex U-Net from \cite{complex_unet} and a real-chanelled counterpart to compare dilation against pooling. The \texttt{complexPyTorch} library \cite{complex-library} was used to implement complex layers as it stores and enumerates real and imaginary weights separately. Accordingly, we increased the filter depths of the real-chanelled models to match the total parameter counts of the complex ones. For further comparisons, we trained DuDorNet, the state-of-the-art multiscale reconstruction network from \cite{dudornet} and the popular D5C5 model from \cite{deepcascade}. To match parameter counts with C5ED, we set the former to use 2 SEDRDB blocks and the latter to have 55 filters per layer. Finally, we also conducted an ablation study removing dilation (dilation rates all set to 1) from C5ED to examine the effectiveness of our multi-scale approach.

\section{Results and Discussion}

\begin{figure}[tp]
    \centering
    \begin{minipage}[t]{\textwidth}
        \setlength{\tabcolsep}{2.5pt}
        \captionof{table}{Average performance of tested methods over each dataset and reduction factor. PSNR (dB), MS-SSIM $\in [0, 1]$.}
        \label{tab:results}
        \tiny
        \sisetup{round-mode=places, round-precision=3}
        \begin{tabularx}{1\textwidth}{l @{\hskip 2em} r l r l r l l l l}
            \toprule
            \multirow{2}{*}[-0.3em]{
            \textbf{Method}}&
            \multicolumn{2}{c}{\textbf{R=4}}&
            \multicolumn{2}{c}{\textbf{R=6}}&
            \multicolumn{2}{c}{\textbf{R=8}}&
            \multirow{2}{*}[-0.3em]{\makecell{\textbf{\#Param}\\\textbf{(M)}}}&
            \strut
            \multirow{2}{*}[-0.3em]{\makecell{\textbf{ms/img}\\\textbf{(inference)}}}&
            \strut
            \multirow{2}{*}[-0.3em]{\makecell{\textbf{min/epoch}\\\textbf{(train)}}}
            
            \\
            
            \cmidrule(lr){2-3} \cmidrule(lr){4-5} \cmidrule(lr){6-7}
        &{PSNR}&{MS-SSIM}&{PSNR}&{MS-SSIM}&{PSNR}&{MS-SSIM}&
            \\\midrule
            
            \multicolumn{10}{c}{\scriptsize \textbf{CC-359 Brains}}

            \\\midrule
            
            \begin{filecontents*}{data/calgary_table.csv}
name,pi,si,pii,sii,piii,siii,params,ms,min
Undersampled,27.803,0.873,23.563,0.759,19.989,0.52,,,
DuDorNet,37.32,0.989,\underline{33.514},0.972,29.206,0.911,0.452,59.551,12.261
$\mathbb{C}$-U-Net,35.262,0.982,31.449,0.954,28.109,0.899,1.1,44.286,7.088
U-Net,35.384,0.982,31.478,0.954,28.4,0.904,1.1,10.985,1.92
D5C5,37.32,0.989,33.025,0.969,28.563,0.903,0.419,7.047,1.963
$\mathbb{C}$-C5ED (Ours),37.391,0.989,33.488,\underline{0.974},29.412,\underline{0.921},0.419,114.388,27.823
C5ED (Ours),\textbf{37.616},\textbf{0.989},\textbf{33.644},\textbf{0.975},\textbf{30.22},\textbf{0.933},0.417,21.184,4.73
Ablation (Ours),\underline{37.409},\underline{0.989},33.361,0.973,\underline{29.474},0.92,0.417,16.901,4.2
\end{filecontents*}
\csvreader[late after line=\\\midrule,late after last line=\\\midrule]{data/calgary_table.csv}{}{\csvcoli & {\csvcolii} & {\csvcoliii} & {\csvcoliv} & {\csvcolv} & {\csvcolvi} & {\csvcolvii} & {\csvcolviii} & {\csvcolix} & {\csvcolx}} 

            \multicolumn{10}{c}{\scriptsize \textbf{FastMRI Knees}}

            \\\midrule

            \begin{filecontents*}{data/fastmri_table.csv}
name,pi,si,pii,sii,piii,siii,params,ms,min
Undersampled,35.414,0.869,31.227,0.803,29.13,0.726,,,
DuDorNet,\textbf{40.389},\textbf{0.904},\textbf{34.887},\textbf{0.858},\textbf{34.374},\textbf{0.845},0.452,104.463,21.167
$\mathbb{C}$-U-Net,39.551,0.9,34.493,0.853,33.507,0.832,1.1,44.913,10.934
U-Net,39.641,0.901,34.497,0.853,33.469,0.832,1.1,12.635,2.67
D5C5,39.914,0.902,32.863,0.851,32.922,0.824,0.419,9.82,3.039
$\mathbb{C}$-C5ED (Ours),\underline{40.317},\underline{0.904},34.662,0.856,33.675,0.837,0.419,152.356,42.422
C5ED (Ours),40.217,0.903,\underline{34.824},\underline{0.858},\underline{34.14},\underline{0.841},0.417,28.682,7.016
Ablation (Ours),40.048,0.902,33.195,0.854,33.815,0.834,0.419,24.464,6.525
\end{filecontents*}
\csvreader[late after line=\\\midrule,late after last line=\\\bottomrule]{data/fastmri_table.csv}{}{\csvcoli & {\csvcolii} & {\csvcoliii} & {\csvcoliv} & {\csvcolv} & {\csvcolvi} & {\csvcolvii} & {\csvcolviii} & {\csvcolix} & {\csvcolx}} 
        \end{tabularx}
    \end{minipage}

    \vspace{2em}
    
    \begin{minipage}[b]{\textwidth}
        \centering
        \includegraphics[width=1\textwidth]{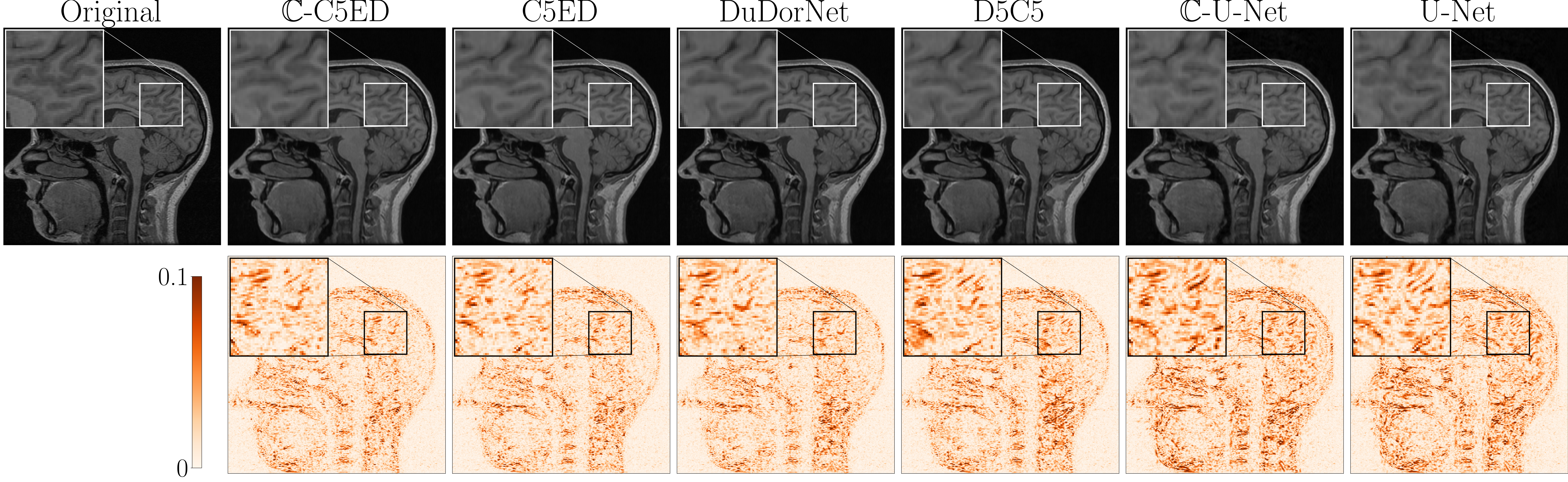}
        \includegraphics[width=1\textwidth]{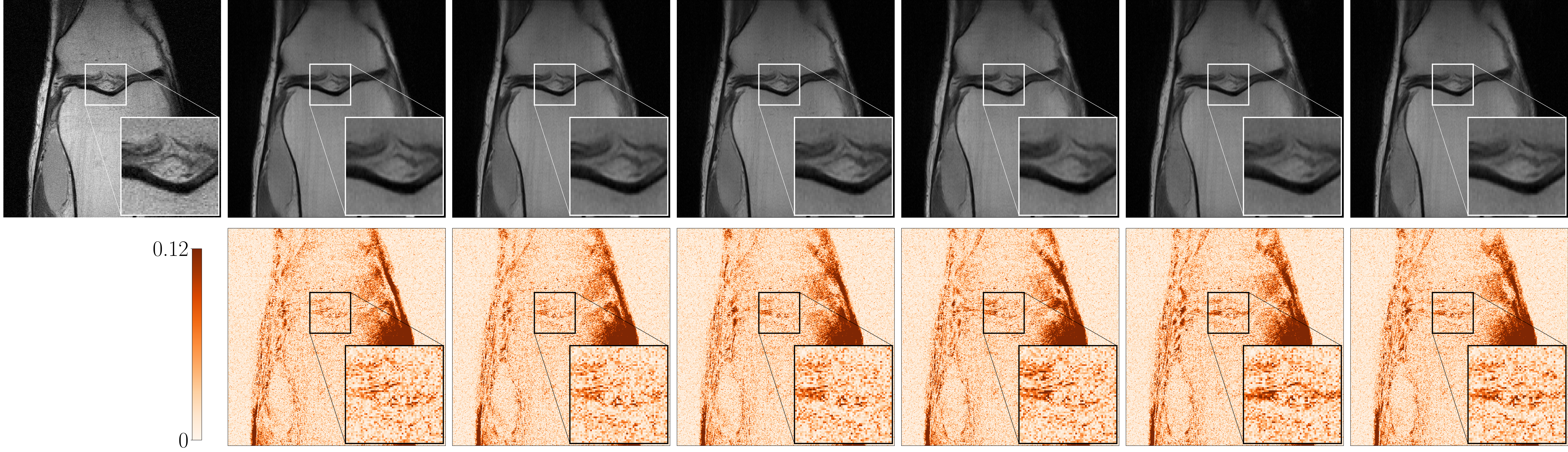}
        \captionof{figure}{Comparison between tested methods on the R6, \sfrac{1}{60} center tiling mask. Our proposed C5-ED and $\mathbb{C}$-C5ED appear visually the sharpest on the brain and knee image respectively. Pixel values $\in [0, 1]$.}
        \label{fig:results}
    \end{minipage}
\end{figure}

Table \ref{tab:results} presents the quantitative metric results for all our experiments and Figure \ref{fig:results} presents example reconstruction outputs for each model.

\noindent\textbf{Complex-valued Results:} Visually, $\mathbb{C}$-C5ED generally produced the best results on the knee dataset whilst it performed similarly to C5ED on the brain dataset. However, the quantitative metrics show the complex-valued models slightly underperformed their real-chanelled equivalents. As the brain images contained primarily magnitude information, it likely did not have sufficient phase information to offer an advantage to complex-valued methods. On the other hand, the knee dataset contained significant background noise which likely obfuscated fine differences in our average-based metrics. This is supported by the MS-SSIM scores, which capture rough multi-scale similarity, being much closer between the real and complex-valued models than PSNR which relies purely on pixel-wise differences that are highly noise sensitive.        

\noindent\textbf{Ablation Study:} The removal of dilation shrank the last two of C5ED's branch RF sizes to $9 \times 9$ (via \eqref{eq:rf_recursive}) from the original's $17 \times 17$ and $35 \times 35$ and reduced performance across the board with larger model separation being observed at higher reduction factors. Qualitatively, the R6 and R8 reconstructions (see supplementary material) resembled the outputs in Figure \ref{fig:learner_outputs} for the smaller branches which reproduced textures well but smeared larger features. These observations suggest C5ED's ability to better restore structural definition is indeed due to its multi-scale approach - by also targeting aliasing at larger feature scales, it can better separate anatomy from structural artefacts too large for the ablation model when undersampling aggressively.

\noindent\textbf{Comparison Results:} C5ED, DuDorNet and D5C5 outperform both U-Nets despite having half as many parameters. Qualitatively, U-Net retains more fine artefacts likely as its resolution bottleneck impedes differentiation between noise and high-frequency detail. Whilst D5C5 doesn't have pooling, neither does it use dilation nor dense connections, so its performance drops noticeably at higher reduction factors with larger artefacts. However, U-Net's downsampling and D5C5's simplicity do offer them significant efficiency advantages, being over six times as fast as DuDorNet. Our model strikes a balance between multi-scale coverage and efficiency - our varied branch depths and ensemble approach cuts down training and inference time by a factor of three over DuDorNet. Moreover, its performance remains highly comparable, surpassing DuDorNet on the brain dataset and only falling slightly short on the knee dataset. This is likely as the brains are much denser in high frequency structures, which better suit the selection of RF sizes and DuDorNet uses $8\%$ more parameters than C5ED.  

\section{Conclusion}
This paper has demonstrated that an ensemble of dilated convolution blocks can effectively reconstruct and de-alias MR images across multiple feature-scales simultaneously. We introduced real and complex-valued versions of a network design featuring a denoising block comprising dilated ensemble branches as well as an overall dense cascade global architecture. Our complex-valued network produced the best qualitative results when phase information was present whilst our real-channel based network outperformed other state-of-the-art complex models. Furthermore, our models matched the performance of a state-of-the-art multi-scale networks whilst offering significant efficiency advantages due to its parallelized architecture.  

\clearpage
\newpage
\bibliography{references.bib}

\input{supplementary}

\end{document}

%% file: supplementary.tex
%


%
\title{Supplementary Material}
\author{}
%
%

\institute{}

\maketitle

\begin{figure}[!h]
\section{Further Implementation Details}
    \begin{minipage}[t]{\textwidth}
        \tiny \centering
        \captionof{table}{Global training settings shared across all experiments.}
        \small
        \begin{tabularx}{1\textwidth}{>{\hsize=.6\hsize}X >{\hsize=0.8\hsize}X >{\hsize=1.6\hsize}X}
            \toprule
            \textbf{Setting} & \textbf{Value} & \textbf{Notes}\\
            \midrule
            GPU & Nvidia Tesla V100 & $1\times$ SXM2 model w/ 32GB VRAM\\
            \midrule
            Epochs & 200 & Checkpoint with least validation loss selected\\
            \midrule
            Learning rate & $5 \cdot 10^{-4}$ & \\
            \midrule
            Batch size & $4$ or $3$ & $4$ for all models except DuDorNet due to VRAM limits\\
            \bottomrule
        \end{tabularx}
        \label{tab:global_params}\\
        \vspace{1em}
        \includegraphics[width=0.4\textwidth]{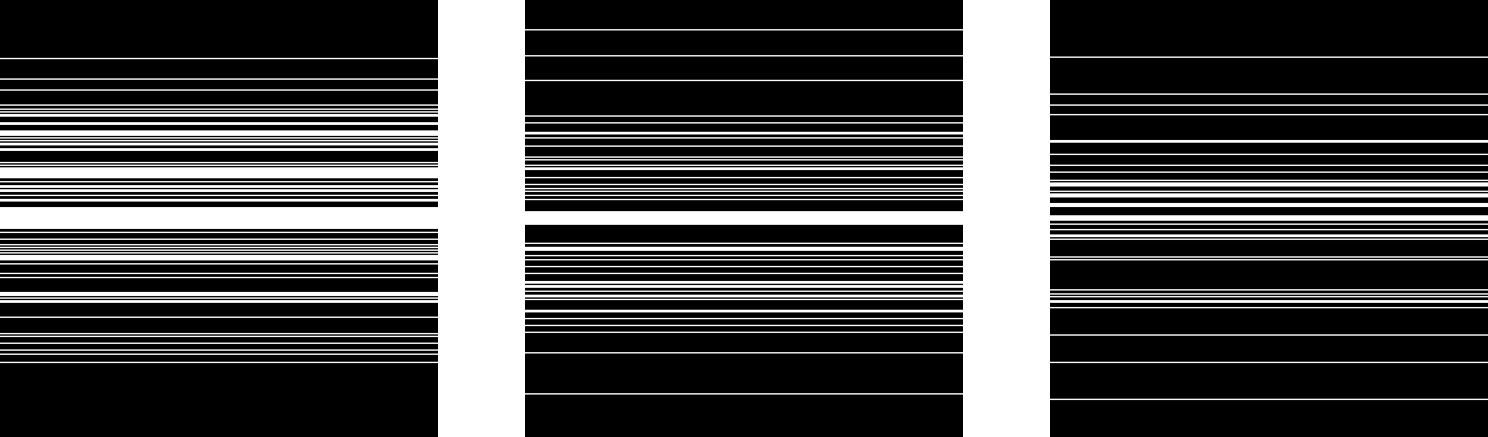}
        \captionof{figure}{R4, R6 and R8 sampling masks, with \sfrac{1}{20}, \sfrac{1}{40} and \sfrac{1}{80} center tiling.}
    \end{minipage}

\end{figure}

\section{Additional Qualitative Results}
\begin{minipage}[!h]{\textwidth}
    \centering
    \includegraphics[width=1\textwidth]{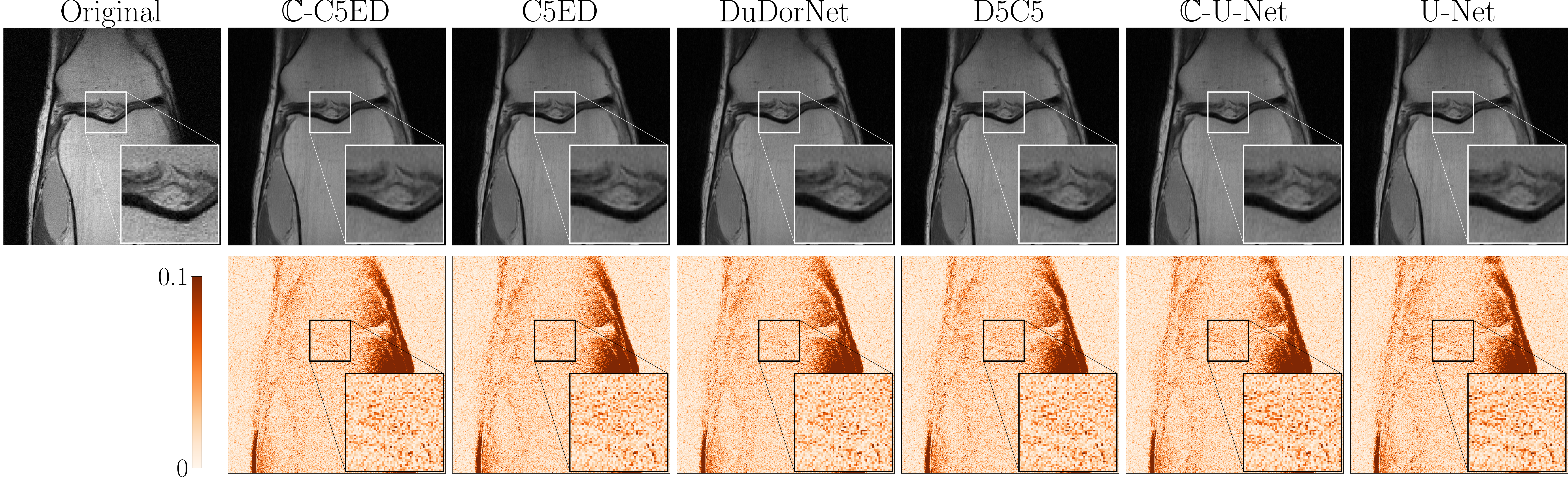}
    \includegraphics[width=1\textwidth]{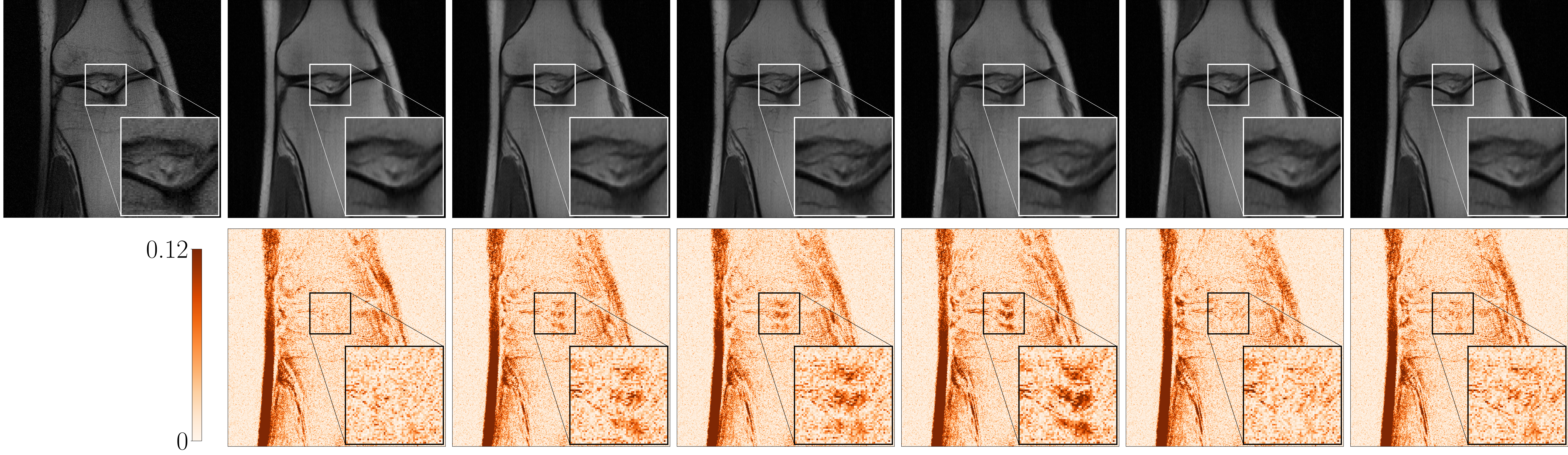}
\end{minipage}
\FloatBarrier

\begin{minipage}[!h]{\textwidth}
    \centering
    \includegraphics[width=1\textwidth, ]{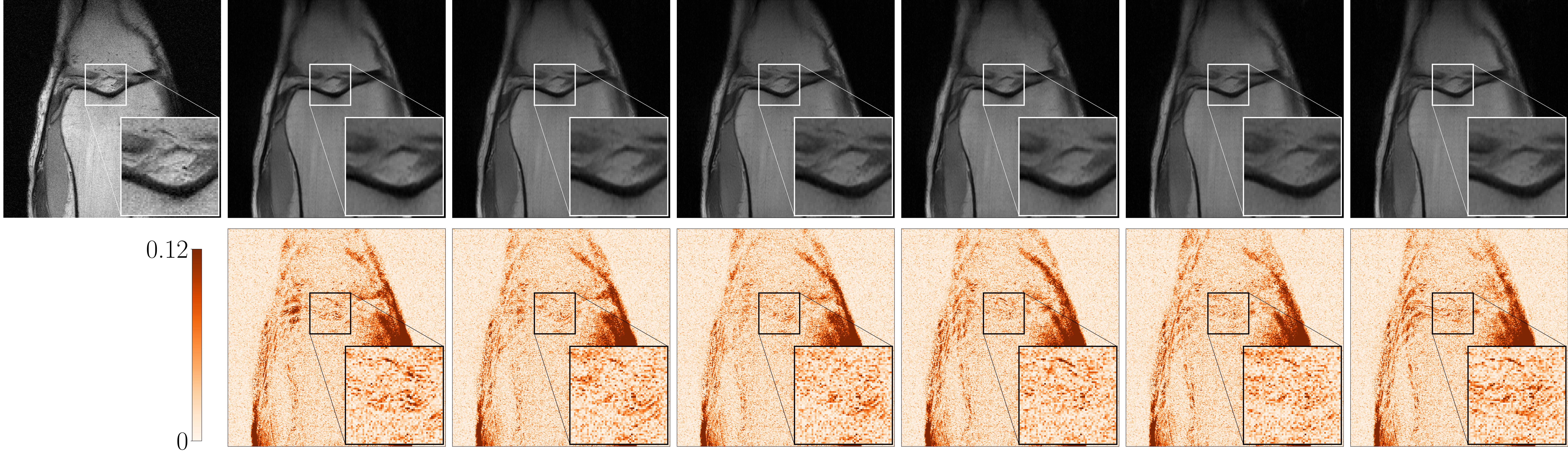}
    \includegraphics[width=1\textwidth]{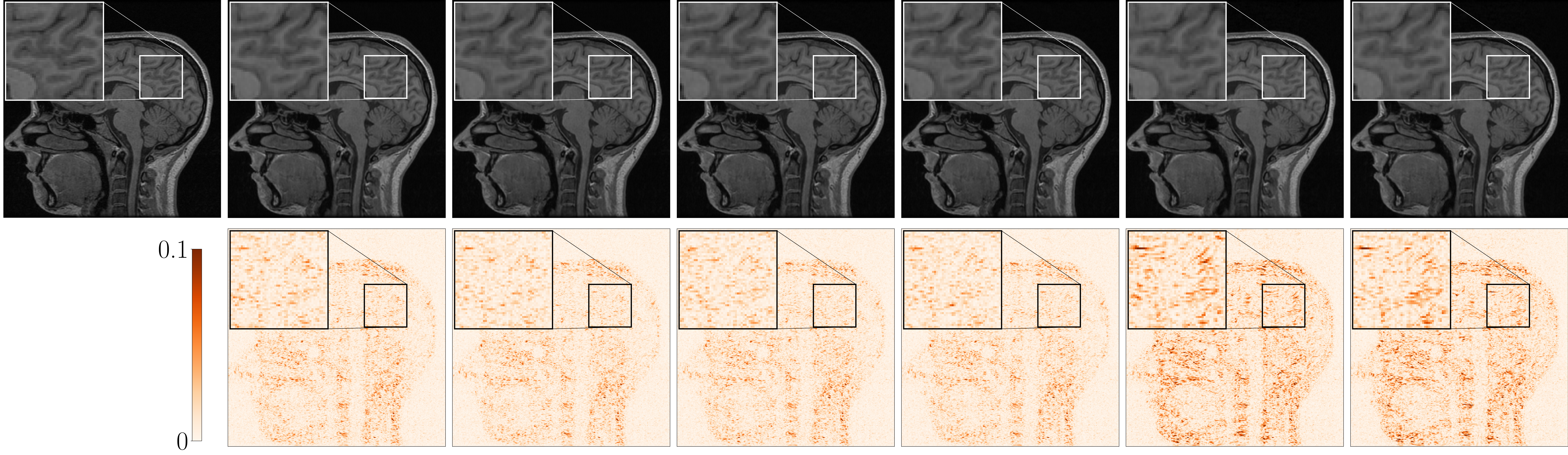}
    \includegraphics[width=1\textwidth]{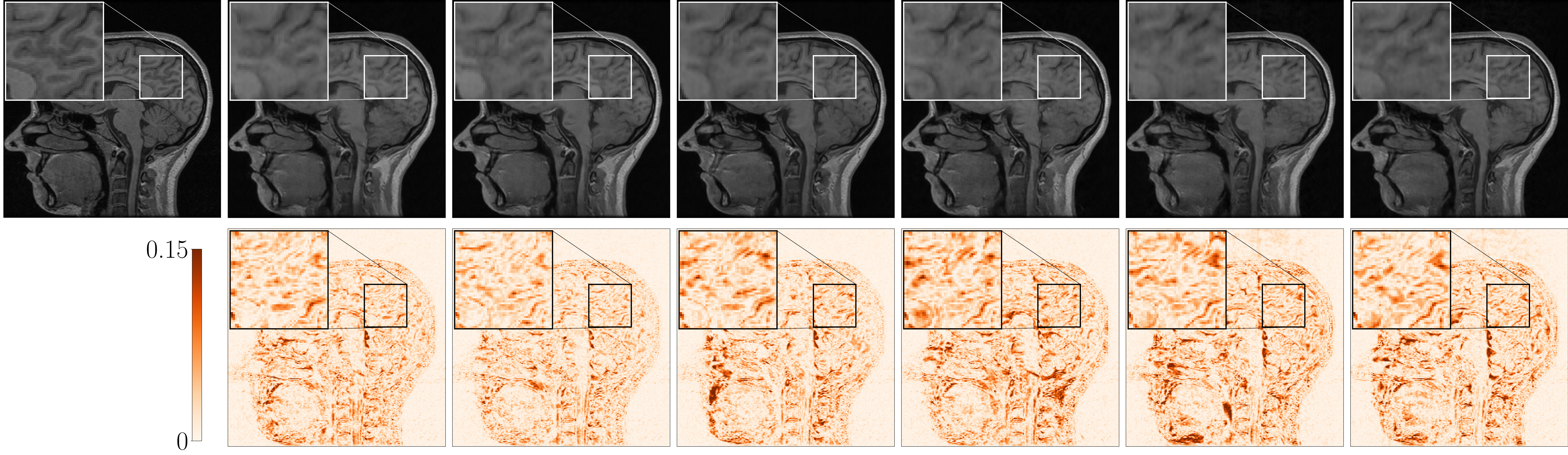}
    \captionof{figure}{Assorted results, reduction factors are R4, R6, R6, R4 and R8.}
    \label{fig:additional_results}

    \vspace{0.5em}

    \includegraphics[width=0.7\textwidth]{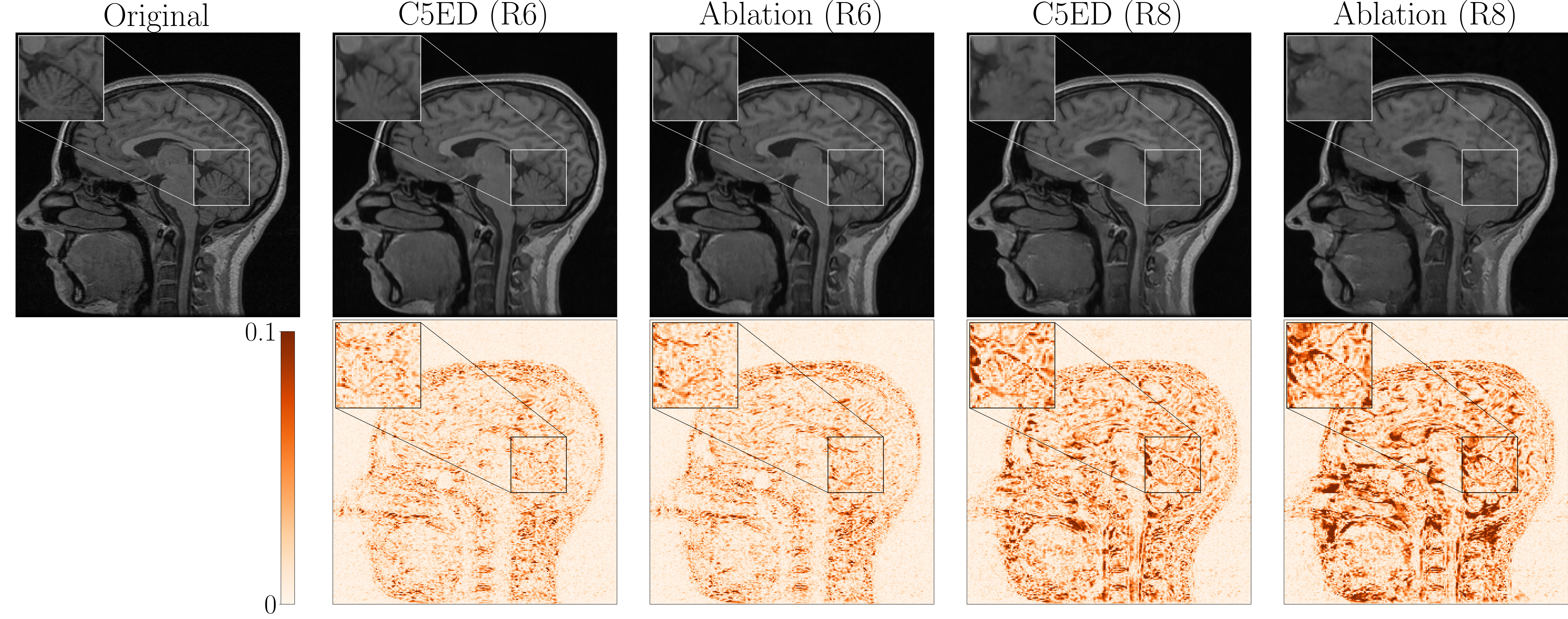}
    \includegraphics[width=0.7\textwidth]{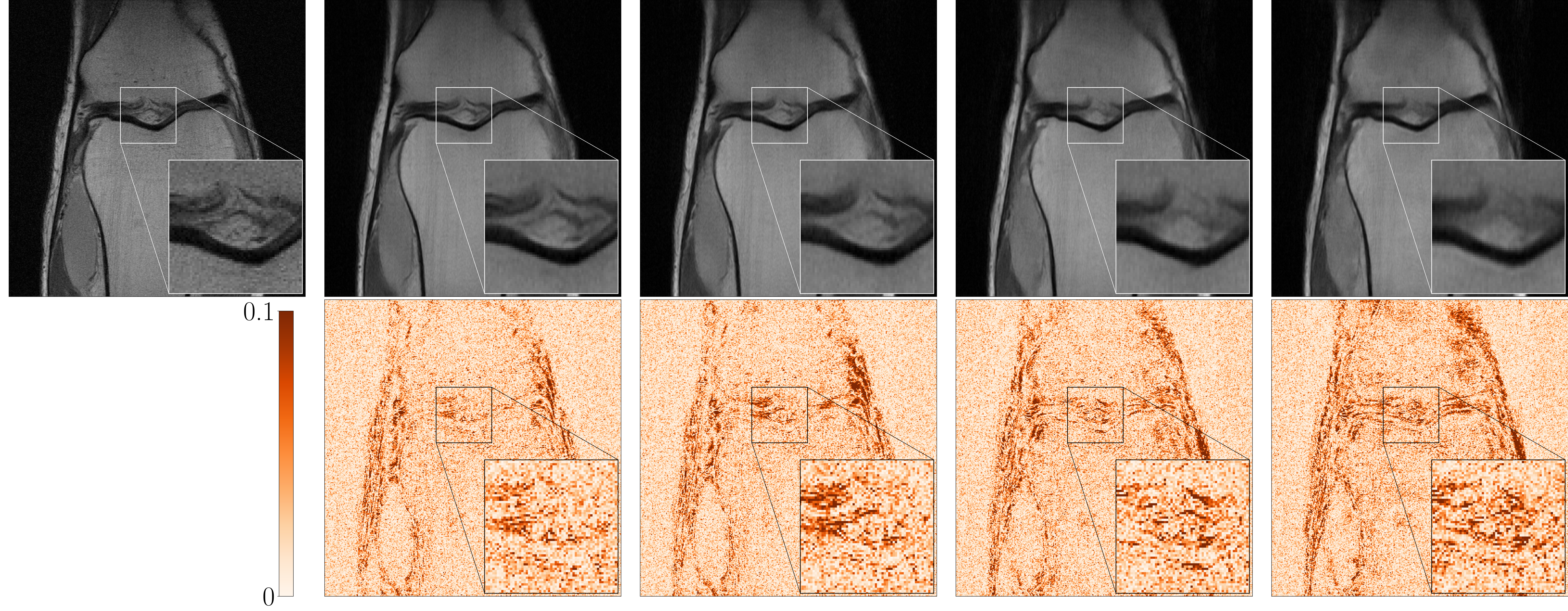}
    \captionof{figure}{C5ED vs. Ablation for R6 and R8 masks. Ablation results are more smeared with less structural definition.}
    \label{fig:ablation_results}
\end{minipage}
\FloatBarrier
